\documentstyle[12pt,graphicx,pstricks,braket,amsmath,amssymb,enumitem,
                    epstopdf,pdfpages,cite,tabu,hyperref,subfigure]{article}
\numberwithin{equation}{section}

\newcommand{\hi}[1]{}

\newcommand{\FbR}{{\bar F}_R}
\newcommand{\Db}{{\bar D}}
\newcommand{\Sb}{{\bar S}}
\newcommand{\phib}{{\bar \phi}}
\newcommand{\FL}{{F}_L}
\newcommand{\FR}{{F}_R}
\newcommand{\oh}{\frac{1}{2}}
\begin{document}

\def\AEF{A.E. Faraggi}

\def\JHEP#1#2#3{{\it JHEP} {\textbf #1}, (#2) #3}
\def\vol#1#2#3{{\bf {#1}} ({#2}) {#3}}
\def\NPB#1#2#3{{\it Nucl.\ Phys.}\/ {\bf B#1} (#2) #3}
\def\PLB#1#2#3{{\it Phys.\ Lett.}\/ {\bf B#1} (#2) #3}
\def\PRD#1#2#3{{\it Phys.\ Rev.}\/ {\bf D#1} (#2) #3}
\def\PRL#1#2#3{{\it Phys.\ Rev.\ Lett.}\/ {\bf #1} (#2) #3}
\def\PRT#1#2#3{{\it Phys.\ Rep.}\/ {\bf#1} (#2) #3}
\def\MODA#1#2#3{{\it Mod.\ Phys.\ Lett.}\/ {\bf A#1} (#2) #3}
\def\RMP#1#2#3{{\it Rev.\ Mod.\ Phys.}\/ {\bf #1} (#2) #3}
\def\IJMP#1#2#3{{\it Int.\ J.\ Mod.\ Phys.}\/ {\bf A#1} (#2) #3}
\def\nuvc#1#2#3{{\it Nuovo Cimento}\/ {\bf #1A} (#2) #3}
\def\RPP#1#2#3{{\it Rept.\ Prog.\ Phys.}\/ {\bf #1} (#2) #3}
\def\APJ#1#2#3{{\it Astrophys.\ J.}\/ {\bf #1} (#2) #3}
\def\APP#1#2#3{{\it Astropart.\ Phys.}\/ {\bf #1} (#2) #3}
\def\EJP#1#2#3{{\it Eur.\ Phys.\ Jour.}\/ {\bf C#1} (#2) #3}
\def\etal{{\it et al\/}}
\def\notE6{{$SO(10)\times U(1)_{\zeta}\not\subset E_6$}}
\def\E6{{$SO(10)\times U(1)_{\zeta}\subset E_6$}}
\def\highgg{{$SU(3)_C\times SU(2)_L \times SU(2)_R \times U(1)_C \times U(1)_{\zeta}$}}
\def\highSO10{{$SU(3)_C\times SU(2)_L \times SU(2)_R \times U(1)_C$}}
\def\lowgg{{$SU(3)_C\times SU(2)_L \times U(1)_Y \times U(1)_{Z^\prime}$}}
\def\SMgg{{$SU(3)_C\times SU(2)_L \times U(1)_Y$}}
\def\Uzprime{{$U(1)_{Z^\prime}$}}
\def\Uzeta{{$U(1)_{\zeta}$}}

\def\lsim{\raise0.3ex\hbox{$\;<$\kern-0.75em\raise-1.1ex\hbox{$\sim\;$}}}
\def\gsim{\raise0.3ex\hbox{$\;>$\kern-0.75em\raise-1.1ex\hbox{$\sim\;$}}}

\newcommand{\cc}[2]{c{#1\atopwithdelims[]#2}}
\newcommand{\bev}{\begin{verbatim}}
\newcommand{\beq}{\begin{equation}}
\newcommand{\bea}{\begin{eqnarray}}
\newcommand{\eea}{\end{eqnarray}}

\newcommand{\beqa}{\begin{eqnarray}}
\newcommand{\beqn}{\begin{eqnarray}}
\newcommand{\eeqn}{\end{eqnarray}}
\newcommand{\eeqa}{\end{eqnarray}}
\newcommand{\eeq}{\end{equation}}
\newcommand{\beqt}{\begin{equation*}}
\newcommand{\eeqt}{\end{equation*}}
\newcommand{\Eev}{\end{verbatim}}
\newcommand{\bec}{\begin{center}}
\newcommand{\eec}{\end{center}}
\newcommand{\bes}{\begin{split}}
\newcommand{\ees}{\end{split}}
\newcommand{\nn}{\nonumber}

\def\ie{{\it i.e.~}}
\def\eg{{\it e.g.~}}
\def\half{{\textstyle{1\over 2}}}
\def\nicefrac#1#2{\hbox{${#1\over #2}$}}
\def\third{{\textstyle {1\over3}}}
\def\quarter{{\textstyle {1\over4}}}
\def\m{{\tt -}}
\def\mass{M_{l^+ l^-}}
\def\p{{\tt +}}

\def\slash#1{#1\hskip-6pt/\hskip6pt}
\def\slk{\slash{k}}
\def\GeV{\,{\rm GeV}}
\def\TeV{\,{\rm TeV}}
\def\y{\,{\rm y}}

\def\l{\langle}
\def\r{\rangle}
\def\LRS{LRS  }

\begin{titlepage}
\samepage{
\setcounter{page}{1}
\rightline{LTH--1128}
\vspace{1.5cm}

\begin{center}
 {\Large \bf 
Wilsonian \\ \medskip Dark Matter in String Derived $Z^\prime$ Model}
\end{center}

\begin{center}

{\large
L. Delle Rose$^\spadesuit$,
A.E. Faraggi$^\clubsuit$,
C. Marzo$^\diamondsuit$
 and
J. Rizos$^\heartsuit$
}\\
\vspace{1cm}
$^\spadesuit${\it School of Physics and Astronomy,  \\
             University of Southampton,
         Southampton SO17 1BJ, UK\\
         Department of Particle Physics, \\ 
         Rutherford Appleton Laboratory,
         Chilton, Didcot, OX11 0QX, UK \\
         }
\vspace{.025in}
$^\clubsuit${\it  Deptartment of Mathematical Sciences,\\
             University of Liverpool,
         Liverpool L69 7ZL, UK\\}
\vspace{.025in}
$^\diamondsuit${\it  National Institute of Chemical Physics and Biophysics, \\
             R\"{a}vala 10, 
             10143 Tallinn,
             Estonia\\}
\vspace{.025in}
$^\heartsuit${\it Department of Physics,
              University of Ioannina, GR45110 Ioannina, Greece \\}

\end{center}

\begin{abstract}

The dark matter issue is among the most perplexing in  
contemporary physics. The problem is more enigmatic due to the 
wide range of possible solutions, ranging from the ultra--light 
to the super--massive.
String theory gives rise to plausible dark matter candidates 
due to the breaking of the non--Abelian Grand Unified Theory (GUT) 
symmetries by Wilson lines. The physical spectrum then contains
states that do not satisfy the quantisation conditions of the 
unbroken GUT symmetry. Given that the Standard Model states 
are identified with broken GUT representations, and provided 
that any ensuing symmetry breakings are induced by components
of GUT states, leaves a remnant discrete symmetry that 
forbid the decay of the Wilsonian states. A class of 
such states are obtained in a heterotic--string derived 
$Z^\prime$ model. The model exploits the spinor--vector duality
symmetry, observed in the fermionic $Z_2\times Z_2$ heterotic--string 
orbifolds, to generate a $Z^\prime\in E_6$ symmetry that may 
remain unbroken down to low energies. The $E_6$ symmetry is broken
at the string level with discrete Wilson lines. The Wilsonian 
dark matter candidates in the string derived model
are $SO(10)$, and hence Standard Model, 
singlets and possess non--$E_6$ $U(1)_{Z^\prime}$ charges. 
Depending on the $U(1)_{Z^\prime}$ breaking scale and the 
reheating temperature they give rise to different scenarios 
for the relic abundance, and in accordance with the
cosmological constraints.

\end{abstract}
\smallskip}
\end{titlepage}

\section{Introduction}
The Standard Model provides viable
parameterisation of all experimental data at the subatomic scale.
Alas, the Standard Model, and point quantum field theories in general, 
is not compatible with the gravitational interaction that accounts 
for observations at the celestial, galactic and cosmological scales. 
Furthermore, the Standard Model contains only a fraction of the 
stable matter required to explain the data at the galactic and
cosmological scales. 

String theory provides
a viable framework for perturbative quantum gravity, and 
gives rise to the gauge and matter states that form the core
of the Standard Model. Furthermore, these ingredients 
arise in string theory due to its internal consistency.
String theory, therefore, provides a consistent framework to 
develop a phenomenological approach to quantum gravity.
The phenomenological heterotic--string models 
constructed in the free fermionic formulation
are among the most realistic string models constructed 
to date. These models are obtained in the vicinity of 
the self--dual point under $T$--duality, providing
plausible symmetry arguments to explain their viability,
and correspond to $Z_2\times Z_2$ toroidal orbifolds,
which are among the most symmetric and simplest 
string compactifications. 

The dark matter conundrum is one of the most perplexing puzzles 
in contemporary observational data. The problem stems from
the plethora of possible solutions and the lack of a clear 
preference for one or the other. Indeed the range of masses 
for potential candidates extend from $10^{59}$GeV in the form
of MACHOS \cite{machos} to $10^{-31}$GeV in the form of ultra--light bosons 
\cite{hotw}. It is prudent therefore to seek guidance from
string theory. In particular, it is sensible to search for 
potential candidates in phenomenological string constructions.

String models contain in them the favoured dark matter 
candidates in the form of stable supersymmetric particles 
and of axion field, 
as well as other dark matter candidates \cite{ssr, hgb, otherdm}.
However, stable supersymmetric dark 
matter requires reliance on a global symmetry, which ordinarily
would be violated in string models \cite{rx}, whereas recent 
observational data seem to disfavour a wide range of 
axion--like candidates \cite{ajeloetal}. Alternative 
dark matter candidates in string vacua exist in the 
form of hidden sector glueball dark matter \cite{hgb}, and
Wilsonian dark matter candidates \cite{ssr}.
The latter category arises in string models due
to the breaking of the non--Abelian GUT (Grand Unified Theory)
gauge symmetries by Wilson lines. 
The physical spectrum in these string models 
contains states that do not satisfy the charge 
quantisation of the unbroken GUT gauge group \cite{ww}.
Specifically, such states carry fractional charge
with respect to some of the unbroken $U(1)$ generators 
of the original GUT symmetry. 
Some of these states may carry fractional electric charge, 
whereas others may carry standard charges under the Standard
Model gauge group, but carry fractional charge 
with respect to an unbroken $U(1)_{Z^\prime}$ gauge symmetry. 
States that carry fractional electric charge are stable 
by virtue of electric charge conservation. States that
carry fractional charge under an unbroken $U(1)_{Z^\prime}$
symmetry may be stable, depending on the charges of the 
states that are used to break the $U(1)_{Z^\prime}$ symmetry. 
Breaking $U(1)_{Z^\prime}$ with Higgs states that carry the 
standard GUT charges under $U(1)_{Z^\prime}$ results in a 
local discrete symmetry that forbids their decay to 
Standard Model states \cite{ssr, lds}. Such states
may therefore be stable and be viable dark matter candidates. 
We dub such states as Wilsonian matter states due to the 
fact that they arise from the breaking of 
non--Abelian GUT symmetries by Wilson lines. 

The possibility of Wilsonian matter states forming the dark matter 
was studied in ref. \cite{ssr} for a variety of possible states,
including fractionally charged states, strongly interacting states
and Standard Model singlet states. The least constrained possibility
takes into account the states that carry standard Standard Model 
charges, but carry fractional charge under a $U(1)_{Z^\prime}$
gauge symmetry. The Wilsonian states investigated in ref. \cite{ssr} 
arise from the symmetry breaking pattern $SO(10)\rightarrow 
SU(3)\times SU(2)\times U(1)^2$. However, the string derived
models that utilise this symmetry breaking pattern
do not contain the required Higgs states with standard 
GUT charges to break the $U(1)_{Z^\prime}$ gauge symmetry
\cite{slm}. Consequently, these models necessarily utilise 
Higgs states with fractional $U(1)_{Z^\prime}$ charges
to break the $U(1)_{Z^\prime}$ along supersymmetric 
flat direction. 
More specifically, the state which 
is missing is the $SU(3)_C\times SU(2)_L\times U(1)_Y$ 
singlet in the $\overline{16}$ representation of $SO(10)$. 
A scan of a large space of similar standard--like vacua 
may reveal the existence of models that do include
the required states \cite{frsslm}. 
However, baring these new constructions, 
the string derived model that we discuss in this paper
provides the first concrete example that realises the 
Wilsonian Standard Model singlet dark matter scenario. 

The models under consideration are
heterotic--string derived models that admit the symmetry
breaking pattern $E_6\rightarrow SO(10)\times U(1)_{\zeta}$, 
with anomaly free $U(1)_\zeta$, in which case $U(1)_\zeta$
can form part of a low scale $U(1)_{Z^\prime}$ combination. 
This is not the case in most of the phenomenological
heterotic--string derived models, in which $U(1)_\zeta$ is
anomalous as a generic consequence of the 
symmetry breaking pattern $E_6\rightarrow SO(10)\times U(1)_\zeta$. 
The construction of the heterotic--string derived model in ref. 
\cite{frzprime} utilises the spinor--vector duality that was 
observed in fermionic $Z_2\times Z_2$ orbifolds \cite{svd1,svd2}. 
The duality operates under the exchange of the total number
of $(16\oplus\overline{16})$ spinorial $SO(10)$ representations
with the total number of vectorial $10$ representations. The models 
that admit an anomaly free $U(1)_\zeta$ gauge symmetry
are self--dual under the spinor--vector
duality. A particular class of models that are self--dual under the
spinor--vector duality map are models in which the $SO(10)\times 
U(1)_\zeta$ symmetry is enhanced to $E_6$. In these models $U(1)_\zeta$
is anomaly free by virtue of its embedding in $E_6$. The total 
number $(16\oplus\overline{16})$ is equal to the total number of 
$10$ representations due to the fact that the $27$ and $\overline{27}$
representations of $E_6$ contain $16+10$ and $\overline{16}+10$,
respectively. Hence, $E_6$ are self--dual under the exchange of the
total number of $(16\oplus\overline{16})$ and the total number of $10$ 
representations. However, 
there exist also self--dual models in which the $SO(10)\times U(1)_\zeta$
gauge symmetry is not enhanced to $E_6$. This is possible 
if the spinorial and vectorial states are obtained from 
different fixed points of the $Z_2\times Z_2$ orbifold.

The string derived $Z^\prime$ model of ref. \cite{frzprime} 
is constructed by 
first selecting a spinor--vector self--dual model at the 
$SO(10)$ level and subsequently breaking the $SO(10)$ symmetry
to the Pati--Salam subgroup. The chiral spectrum of the resulting
Pati--Salam string model respects the self--duality under the 
spinor--vector duality. Effectively, the result is that 
the chiral spectrum forms complete $E_6$ multiplets and consequently
$U(1)_\zeta$ is anomaly free. 

An unexpected result that was obtained in ref. \cite{frzprime} is with respect to the type
of exotic states that appear in the model. Using 
the trawling algorithm developed for the classification
of free fermionic $Z_2\times Z_2$ orbifolds 
\cite{gkr,fknr,psclass, fsu5class, su62}, we fish out
a model in which all the exotic fractionally charged 
states are projected out from the massless spectrum,
and appear only as massive states \cite{psclass}. 
Such models are dubbed exophobic string models. Therefore,
the model does not contain any massless states with 
fractional charges with respect to the $SO(10)$ subgroup. 
However, the model contains exotic states with respect 
to the $E_6$ subgroup, {\it i.e.} the model contains
states that are $SO(10)$ singlets 
and carry fractional non--$E_6$ charge under $U(1)_\zeta$. 
It is noted that as the gauge symmetries are realised
in this model, as level one Kac--Moody algebras the aforementioned
exotic charges cannot arise from higher order $E_6$ 
representations. Furthermore, the model does contain
the required standard $E_6$ states to break $U(1)_{\zeta}$ 
along flat directions. The model of ref \cite{frzprime} 
therefore, and for the first time, precisely realises 
the Wilsonian Standard Model singlet dark matter scenario
alluded to in ref. \cite{ssr}. 

Our paper is organised as follows: in section \ref{wilsonian}
we discuss and classify the type of exotic states that 
arise in the phenomenological fermionic $Z_2\times Z_2$ 
models. We discuss the structure of the models and their 
construction. In section \ref{e6wilstate} we elaborate on
the exotic $E_6$ states that are obtained in the $Z^\prime$ 
model of ref. \cite{frzprime}.
In section \ref{wildarmat} we investigate 
the exotic $E_6$ states as dark matter candidates, taking 
into account low and high scale $U(1)_{Z^\prime}$ breaking
as well as scenarios with and without inflation. 
Section \ref{concusion} concludes our paper. 

\section{Wilsonian states}\label{wilsonian}

The class of string models under consideration are constructed
in the free fermionic formulation \cite{fff}.
The four dimensional heterotic--string in the light--cone 
gauge requires 20 left--moving, and 44 right--moving,
real fermions propagating on the string worldsheet. 
The sixty--four worldsheet fermions are typically denoted by 
$\{
\psi^{1,2}, 
(\chi, y, \omega)^{1,\cdots,6}\vert
({\bar y}, {\bar\omega} )^{1,\cdots,6},
{\bar\psi}^{1,\cdots,5}, 
{\bar\eta}^{1,2,3},
{\bar\phi}^{1,\cdots,8}
\}
$,
where 32 of the right--moving real fermions 
are grouped into 16 complex fermions that 
produce the Cartan generators of a rank 16 gauge group. 
Here ${\bar\psi}^{1,\cdots,5}$ are the 
Cartan generators of the $SO(10)$ GUT group and
${\bar\phi}^{1,\cdots,8}$ are the Cartan generators
of the rank eight hidden sector gauge group. 
The three complex worldsheet fermions 
${\bar\eta}^{1,2,3}$ generate three Abelian
currents, $U(1)_{1,2,3}$,
in the Cartan subalgebra of the four dimensional gauge 
group with $U(1)_\zeta$ being their linear combination
\beq
U(1)_\zeta ~=~ U(1)_1+U(1)_2+U(1)_3~.
\label{u1zeta}
\eeq
The worldsheet fermions 
pick up a phase under parallel transport around 
one of the non--contractible loops of the vacuum to
vacuum torus amplitude.
These phases are encoded in forty--eight dimensional vectors
$(20_{l.r.}+12_{r.r.}+32_{r.c.})$, 
$$v=\{v(f_1), \cdots, v(f_{20})\vert
v({\bar f}_1), \cdots, v({\bar f}_{48})\}.$$
Invariance under modular 
transformations of the one--loop vacuum to vacuum amplitude
leads to a set of constrains on the phase assignments.
Summation over all the allowed phases, with appropriate 
phases to render the sum modular invariant, generates
the partition function. The string vacua in the free 
fermionic formulation are obtained by specifying 
a set of boundary condition basis vectors,
$B=\{v_1, v_2, v_3, \cdots\}$, and 
the one--loop summation phases in the partition function
$\cc{v_i}{v_j}$. 
The basis set spans a space $\Xi$, which consists of 
all possible linear combinations of the basis vectors
$\Xi= \sum_kn_kv_k$, where $n_k=0, \cdots, N_{v_k} -1$, and $N_{v_k}$ 
denote the order of each of the basis vectors. The physical 
states in the Hilbert space of a given sector $\xi\in\Xi$ are 
obtained by acting on the vacuum with fermionic and bosonic 
oscillators and by imposing the Generalised GSO (GGSO) projections. 
The $U(1)$ charges with respect to the Cartan generators of the 
four dimensional gauge group are given by
$$Q(f) = \frac{1}{2}\xi(f) + F_\xi(f), $$
where $\xi(f)$ is the boundary condition of the complex 
worldsheet fermion $f$ in the sector $\xi$, and $F_\xi(f)$ is 
a fermion number operator \cite{fff}.
The phenomenological properties of the models are 
extracted by calculating tree-level and higher order terms in the 
superpotential and by analysing its flat directions. It is important
to note that the free fermionic models correspond to toroidal 
$Z_2\times Z_2$ orbifolds at special points in the moduli space
\cite{z2z2corres}. Moduli deformations of the six dimensional 
internal torus are incorporated in the fermionic construction 
in terms of worldsheet Thirring interactions that are consistent
with the transformation properties of the worldsheet fermions. 

Early examples of quasi--realistic free fermionic models 
corresponded to the so--called NAHE--based models. The first
set of five basis vectors, dubbed the NAHE--set \cite{nahe}, 
is common in all these phenomenological models, and the models
vary by the addition of three or four basis vectors beyond the 
NAHE--set.
Three generation models with $SU(5)\times U(1)$ (FSU5) 
\cite{fsu5}; $SO(6)\times SO(4)$ (SO64) \cite{PSmodels}; 
$SU(3)\times SU(2)\times U(1)^2$  (SLM) \cite{slm}; and
$SU(3)\times U(1)\times SU(2)^2$ (LRS) \cite{lrs} $SO(10)$
subgroup were obtained, whereas the case with 
$SU(4)\times SU(2)\times U(1)$ was shown not to produce
viable models \cite{su421}. 
In more recent years systematic methods were developed 
for the classification of large spaces of fermionic 
$Z_2\times Z_2$ heterotic--string vacua. The classification
methodology uses an appropriate fixed set of boundary
condition basis vectors and the space of vacua 
is spanned by varying the GGSO projection coefficients. 
In this manner models with unbroken $SO(10)$ gauge group 
were classified \cite{fknr},
which led to the observation of the spinor--vector duality
\cite{svd1}, as well as models with SO64 \cite{psclass} and 
FSU5 \cite{fsu5class} $SO(10)$ subgroups.
Classification of SLM and LRS models is currently
underway and will be reported in future publications. 
 
The construction of free fermionic models, in either 
the older trial and error method, or the more recent systematic 
classification method, can be viewed in two stages. The first 
part consist of the basis vectors that 
preserve the $SO(10)$ symmetry. The construction at 
this stage produces vacua with $(2,0)$ worldsheet supersymmetry,
$N=1$ spacetime supersymmetry, and a number of 
spinorial and vectorial representations of $SO(10)$.
The second part consist of the inclusion of the basis
vectors that break the $SO(10)$ symmetry to a subgroup. 

Correspondingly, the sectors in a free fermionic heterotic--string 
model can be divided into those that preserve the $SO(10)$ symmetry
and those that do not. Physical states that arise from sectors 
that preserve the $SO(10)$ symmetry correspond to states 
that may be identified with Standard Model states, or
are Standard Model singlets. Sectors that break the 
$SO(10)$ symmetry produce exotic states, {\it i.e.} 
they produce states that carry fractional charge 
under $U(1)_{\rm e.m.}$ or under $U(1)_{Z^\prime}\in SO(10)$. 
The exotic states can be further classified 
according to the $SO(10)$ symmetry breaking
pattern in the sector from which they arise.

The Cartan subalgebra of the observable gauge group in
the free fermionic models is generated by the complex worldsheet 
fermions $\{{\bar\psi}^{1,\cdots,5}, {\bar\eta}^{1,2,3}\}$, 
with ${\bar\psi}^{1,\cdots,5}$ producing those of $SO(10)$ 
and its subgroups and ${\bar\eta}^{1,2,3}$ producing three 
$U(1)$ currents. The $SO(10)$ symmetry is broken to 
one of its subgroups by the following assignments:
\beqn
&1.&b\{{{\bar\psi}_{1\over2}^{1\cdots5}}\}=
\{{1\over2}{1\over2}{1\over2}{1\over2}
        {1\over2}\}\Rightarrow SU(5)\times U(1),\label{su51breakingbc}\\
&2.&b\{{{\bar\psi}_{1\over2}^{1\cdots5}}\}=\{1~ 1\, 1\, 0\, 0\}\,
  \Rightarrow SO(6)\times SO(4).
\label{so64breakingbc}
\eeqn
To break the $SO(10)$ symmetry to
$SU(3)_C\times SU(2)_L\times
U(1)_C\times U(1)_L$ \cite{slm}
both steps, 1 and 2, are used, in two separate basis
vectors\footnote{$U(1)_C={3\over2}U(1)_{B-L};
U(1)_L=2U(1)_{T_{3_R}}.$}.
The breaking pattern
$SO(10)\rightarrow SU(3)_C\times SU(2)_L\times SU(2)_R \times U(1)_{B-L}$
\cite{lrs} is obtained with: 
\beqn
&3.&b\{{{\bar\psi}_{1\over2}^{1\cdots5}}\}=
\{{1\over2}{1\over2}{1\over2}00\}\Rightarrow SU(3)_C\times U(1)_C
\times SU(2)_L\times SU(2)_R,
\label{su3122breakingbc}
\eeqn
and the breaking pattern
$SO(10)\rightarrow SU(4)_C\times SU(2)_L\times U(1)_R$ \cite{su421}
results from: 
\beqn
&4.& b\{{{\bar\psi}_{1\over2}^{1\cdots5}}\}=
\{000{1\over2}{1\over2}\}\Rightarrow SU(4)_C\times
SU(2)_L\times U(1)_R.
\label{su421breakingbc}
\eeqn
It was shown
that the breaking pattern
(\ref{su421breakingbc}) does not produce viable models \cite{su421}.

The states in the free fermionic models that carry exotic charges with respect
the Abelian generators of the $SO(10)$ subgroups can be classified according to 
the $SO(10)$ symmetry breaking pattern in the sectors from which they arise. 
A basis vector combination that produces exotic states
contains in it the $SO(10)$ breaking basis vectors. 
We focus here on the case of the standard--like models that 
contain both of the assignments shown in (\ref{su51breakingbc})
and in (\ref{so64breakingbc}) and therefore contain
the exotic states that arise in the FSU5 and the SO64 models, 
as well as those that arise solely in the SLM models. 
In the following we adapt the notation 
\begin{equation}
[(SU(3)_C\times U(1)_C);
     (SU(2)_L\times U(1)_L)]_{(Q_Y,Q_{Z^\prime},Q_{\rm e.m.})},
\label{notation}
\end{equation}
to denote the charges of the states arising in the exotic sectors. 
Here $U(1)_C$ and $U(1)_L$ are defined in terms of the 
worldsheet charges by
\beq
Q_C=Q({\bar\psi}^1)+Q({\bar\psi}^2)+ Q({\bar\psi}^3)~
{\rm and } ~
Q_L=Q({\bar\psi}^4)+Q({\bar\psi}^5).
\label{qcql}
\eeq
The FSU5 $U(1)$ combinations are given by
\beqn
U(1)_5 & =  & \frac{1}{3}U(1)_C ~-~ \frac{1}{2} U(1)_L ~~\in~~ SU(5)\\
~~~ 
U(1)_{\tilde 5} & = & ~~U(1)_C ~+~  ~~U(1)_L ~~\notin~~ SU(5)
\eeqn
whereas the weak hypercharge and ${Z^\prime}$ charges in eq. 
(\ref{notation}) are given by
\beqn
U(1)_Y & =  & \frac{1}{3}U(1)_C ~+~ \frac{1}{2} U(1)_L ~~=~~ 
                \frac{2}{5}U(1)_5 ~-~\frac{1}{5}U(1)_{\tilde 5}
\label{u1y}
\\
~~~ 
U(1)_{Z^\prime} & = & ~~U(1)_C ~-~  ~~U(1)_L ~~=~~
                \frac{12}{5}U(1)_5 ~+~\frac{1}{5}U(1)_{\tilde 5}  .
\label{u1zprime}       
\eeqn
$U(1)_C$ and $U(1)_L$ are similarly defined in the SO64 models. The 
electromagnetic charge is given by
\beq
U(1)_{\rm e.m.} ~=~ T_{3_L}~+~ U(1)_Y .
\eeq
Using the notation in eq. (\ref{notation}) the Standard Model 
matter states carry the charges
\beqn
{e_L^c}&\equiv& ~~[(1,{3\over2});(1,1)]_{(1,1/2,1)};\label{elc}\\
{u_L^c}&\equiv& ~~[({\bar 3},-{1\over2});(1,-1)]_{(-2/3,1/2,-2/3)};
							\label{ulc}\\
Q&\equiv& ~~[(3,{1\over2});(2,0)]_{(1/6,1/2,(2/3,-1/3))};\label{q}\\
{N_L^c}&\equiv& ~~[(1,{3\over2});(1,-1)]_{(0,5/2,0)};\label{nlc}\\
{d_L^c}&\equiv& ~~[({\bar 3},-{1\over2});(1,1)]_{(1/3,-3/2,1/3)};
							\label{dlc}\\
L&\equiv& ~~[(1,-{3\over2});(2,0)]_{(-1/2,-3/2,(0,1))},\label{l}
\eeqn
and arise from spinorial 16 representations of $SO(10)$, {\it i.e.}
they arise from sectors that preserve the $SO(10)$ gauge symmetry.
Similarly, the light Higgs electroweak doublets are obtained from $SO(10)$ 
vectorial representations, and arise in sectors that preserve the 
$SO(10)$ symmetry. By contrast the exotic states arise in sectors
that break the $SO(10)$ symmetry and can be classified according to the 
$SO(10)$ symmetry breaking pattern in each sector. 
Sectors that break the $SO(10)$ symmetry to the FSU5 subgroup contain
in them the assignment in eq. (\ref{su51breakingbc}) and produce the states
\beqn
& &[(3,-{1\over4});(1,~~{1\over2})]_{(1/6,-3/4,1/6)}~~~~;~~~~
  [(\bar3,~~{1\over4});(1,-{1\over2})]_{(-1/6,~~3/4,-1/6)},
\label{fsu5etrip} \\
& &[(1,~~{3\over4});(2,-{1\over2})]_{(~0~,5/4,\pm1/2)}~~~~;~~~~
  [(1,-{3\over4});(2,~~{1\over2})]_{(~~~0~,-5/4,\pm1/2)},
\label{fsu5edoub} \\
& &[(1,~~{3\over4});(1,~~{1\over2})]_{(1/2,1/4,~~1/2)}~~~~;~~~~
  [(1,-{3\over4});(1,-{1\over2})]_{(-1/2,-1/4,-1/2)}. 
\label{fsu5esing}
\eeqn
Sectors that break the $SO(10)$ symmetry to the SO64 subgroup contain
in them the assignment in eq. (\ref{so64breakingbc}) and produce the states
\beqn
& &[(3,~~{1\over2});(1,~{~0})]_{(1/6,~~1/2,~~1/6)}~~~~;~~~~
  [(\bar3,-{1\over2});(1,~{~0})]_{(-1/6,-1/2,-1/6)},
\label{so64etrip} \\
& &[(1,~~{3\over2});(1,~{~0})]_{(1/2,~~3/2,~~1/2)}~~~~;~~~~
  [(1,-{3\over2});(1,~~{~0})]_{(-1/2,-3/2,-1/2)},
\label{so64esing1} \\
& &[(1,~{~0});(1,~~{~1})]_{(1/2,~~-1,~~1/2)}~~~~;~~~~
   [(1,~{~0});(1,~-{ 1})]_{(-1/2,~~1~,-1/2)},
\label{so64esing2}\\
& &[(1,~{~0});(2,~~{~0})]_{(~~0,~~~0,~\pm1/2)}~~~~.
\label{so64edoub}
\eeqn
Sectors that break the $SO(10)$ symmetry to the SLM subgroup contain 
a linear combination of both assignments in eq.(\ref{su51breakingbc})
and eq. (\ref{so64breakingbc}). These sectors produce states that carry 
standard GUT charges with respect to the Standard Model gauge group
but carry fractional non--GUT charges with respect to the $U(1)_{Z^\prime}$ 
combination in eq. (\ref{u1zprime}): 
\beqn
& &[(3,-{1\over4});(1,-{1\over2})]_{(-1/3,~~1/4,-1/3)}~~~~;~~~~
   [(\bar3,~~{1\over4});(1,~~{1\over2})]_{(~~1/3,-1/4,~~1/3)},
\label{slmetrip} \\
& &[(1,~~{3\over4});(2,-{1\over2})]_{(~~1/2,~~3/4,~(1,0))}~~~~;~~~~
  [(1,-{3\over4});(2,~~{1\over2})]_{(-1/2,-3/4,(0,-1))},~~
\label{slmedoub} \\
& &[(1,~~{3\over4});(1, -{1\over2})]_{(~~~0~,~~5/4,~~~0~)}~~~~;~~~~
  [(1,-{3\over4});(1,-{1\over2})]_{(~~~0,-5/4,~~~0)}. 
\label{slmesing}
\eeqn

The exotic states appearing in eqs. 
(\ref{slmetrip},\ref{slmedoub}, \ref{slmesing})
may therefore produce viable dark matter candidates. This would be
the case if the heavy Higgs states that break $U(1)_{Z^\prime}$
carry the standard GUT charges with respect to $U(1)_{Z^\prime}$. 
In that case a remnant discrete symmetry forbids the formation of
unsuppressed terms that can lead to decay of the exotic states to 
the Standard Model states \cite{ssr}. 
We remark that all nonrenormalisable gauge invariant 
operators that may be formed are suppressed by at least
one power of $M_{\rm String}$. They are therefore sufficiently small
and cannot lead to rapid decay of the Wilsonian dark matter
states \cite{ssr}.  
In the FSU5 and SO64 heavy Higgs states necessarily arise 
from GUT representation in order to break the remaining 
non--Abelian symmetry. However, in the SLM models this need 
not be the case. The remnant unbroken symmetry 
$U(1)_{Z^\prime}$ of eq. (\ref{u1zprime}) can be broken by
heavy Higgs states with the standard GUT 
charges of eq. (\ref{nlc}) and its conjugate ${\bar N}_L^c$, 
or by using the exotic states and charges in eq. (\ref{slmesing}). 
In the SLM models of \cite{slm}, 
a state with the quantum numbers of ${\bar N}_L^c$ 
does not appear in the massless spectrum. 
Consequently, in these models, 
breaking $U(1)_{Z^\prime}$ and 
preserving supersymmetry at a high scale forces the exotic
states in eq. (\ref{slmesing}) to get a non--trivial VEV 
at the high scale. 
Alternatively, we may contemplate that 
either the $U(1)_{Z^\prime}$ symmetry is broken at the low scale, 
or that supersymmetry is broken at the high scale. 
Suppression of 
left--handed neutrino masses by the seesaw mechanism 
disfavours the first possibility. 
Breaking supersymmetry at the high scale also introduces a 
plethora of new questions that we do not consider in this paper. 
The upshot is that the Wilsonian singlet dark matter scenario of \cite{ssr},
with the type singlets in eq. (\ref{slmesing}),
is not realised in the existing SLM heterotic--string free fermionic
models.

\section{$E_6$ Wilsonian states in $Z^\prime$ string model }
\label{e6wilstate}

We next turn to discuss the Wilsonian matter states in the string 
derived model of ref. \cite{frzprime}. The $SO(10)$ symmetry
is broken in this model to the Pati--Salam subgroup. 
The chiral spectrum of the model forms complete $E_6$ multiplets. 
Consequently, the $U(1)_\zeta$ combination, which possesses
the embedding $SO(10)\times U(1)_\zeta\in E_6$ is anomaly free. 
The complete massless spectrum of the model, and its charges 
under the four dimensional gauge group, is given in ref. \cite{frzprime}. 
Here a glossary of the states in the model and the charges under
the $SU(4)\times SO(4)\times U(1)_\zeta$ are shown in
tables \ref{tableb} and \ref{tablehi}.
\begin{table}[!h]
\begin{tabular}{|c|c|c|c|}
\hline
Symbol& Fields in \cite{frzprime} & 
                         $SU(4)\times{SU(2)}_L\times{SU(2)}_R$&${U(1)}_{\zeta}$\\
\hline
$\FL$ &        $F_{1L},F_{2L},F_{3L}$&$\left({\bf4},{\bf2},{\bf1}\right)$&$+\oh$\\
$\FR$ &$F_{1R}$&$\left({\bf4},{\bf1},{\bf2}\right)$&$-\oh$\\
$\FbR$&${\bar F}_{1R},{\bar F}_{2R},{\bar F}_{3R},{\bar F}_{4R}$
                             &$\left({\bf\bar 4},{\bf1},{\bf2}\right)$&$+\oh$\\
$h$   &$h_1,h_2,h_3$&$\left({\bf1},{\bf2},{\bf2}\right)$&$-1$\\
$\Delta$&$ D_1,\dots,  D_7$&$\left({\bf6},{\bf1},{\bf1}\right)$&$-1$\\
$\bar\Delta$&$\Db_1,\Db_2,\Db_3,\Db_6$&$\left({\bf6},{\bf1},{\bf1}\right)$&$+1$\\
$S$&$\Phi_{12},\Phi_{13},\Phi_{23},\chi^+_1,\chi^+_2,\chi^+_3,\chi^+_5$
&$\left({\bf1},{\bf1},{\bf1}\right)$&$+2$\\
$\Sb$&$\bar\Phi_{12},\bar\Phi_{13},\bar\Phi_{23},\bar\chi^+_4$&$\left({\bf1},{\bf1},{\bf1}\right)$&$-2$\\
$\phi$&$\phi_1,\phi_2$&$\left({\bf1},{\bf1},{\bf1}\right)$&$+1$\\
$\phib$&$\bar\phi_1,\bar\phi_2$&$\left({\bf1},{\bf1},{\bf1}\right)$&$-1$\\
$\zeta$&$\Phi_{12}^-,\Phi_{13}^-,\Phi_{23}^-,\bar\Phi_{12}^-,\bar\Phi_{13}^-,\bar\Phi_{23}^-$&$\left({\bf1},{\bf1},{\bf1}\right)$&$\hphantom{+}0$\\
&$\chi_1^-,\chi_2^-,\chi_3^-,\bar\chi_4^-,\chi_5^-$&$$&$$\\
&$\zeta_i,\bar\zeta_i,i=1,\dots,9$&$$&$$\\
&$\Phi_i,i=1,\dots,6$&$$&$$\\
\hline
\end{tabular}
\caption{\label{tableb}
Observable sector field notation and associated states in \cite{frzprime}.}
\end{table}
\begin{table}[!h]
\begin{center}
\begin{tabular}{|c|c|c|c|}
\hline
Symbol& Fields in \cite{frzprime} & ${SU(2)}^4\times SO(8)$&${U(1)}_{\zeta}$\\
\hline
$H^+$&$H_{12}^3$&$\left({\bf2},{\bf2},{\bf1},{\bf1},{\bf1}\right)$&$+1$\\
&$H_{34}^2$&$\left({\bf1},{\bf1},{\bf2},{\bf2},{\bf1}\right)$&$+1$\\
$H^-$&$H_{12}^2$&$\left({\bf2},{\bf2},{\bf1},{\bf1},{\bf1}\right)$&$-1$\\
&$H_{34}^3$&$\left({\bf1},{\bf1},{\bf2},{\bf2},{\bf1}\right)$&$-1$\\
$H$&$H_{12}^1$&$\left({\bf2},{\bf2},{\bf1},{\bf1},{\bf1}\right)$&$0$\\
&$H_{13}^i,i=1,2,3$&$\left({\bf2},{\bf1},{\bf2},{\bf1},{\bf1}\right)$&$0$\\
&$H_{14}^i,i=1,2,3$&$\left({\bf2},{\bf1},{\bf1},{\bf2},{\bf1}\right)$&$0$\\
&$H_{23}^1$&$\left({\bf1},{\bf2},{\bf2},{\bf1},{\bf1}\right)$&$0$\\
&$H_{24}^1$&$\left({\bf1},{\bf2},{\bf1},{\bf2},{\bf1}\right)$&$0$\\
&$H_{34}^i,i=1,4,5$&$\left({\bf1},{\bf1},{\bf2},{\bf2},{\bf1}\right)$&$0$\\
$Z$&$Z_i,i=1,\dots,$&$\left({\bf1},{\bf1},{\bf8}\right)$&$0$\\
\hline
\end{tabular}
\end{center}
\caption{\label{tablehi}
Hidden sector field notation and associated states in \cite{frzprime}. }
\end{table}
The heterotic--string model in ref. \cite{frzprime} is an exophobic 
Pati--Salam model. The type of massless exotic states that can appear 
in this model are those in eqs. (\ref{so64etrip}, \ref{so64edoub},
\ref{so64esing1}, \ref{so64esing2}). 
However, none of these states appear in the massless spectrum. 
In fact, none of the exotic states discussed in section \ref{wilsonian}
appear in this model. 

The model contains, however, a new type of exotic states. 
These exotic states carry standard $SO(10)$ charges
and are in fact $SO(10)$ singlets. They are exotic with respect to 
$U(1)_\zeta$ as they carry 1/2 of the charge of the standard 
$SO(10)$ singlets in the 27 and $\overline{27}$ representations
of $E_6$. Inspection of table \ref{tableb} shows that the 
exotic states of this type are $\{ \phi_{1,2}, {\bar\phi}_{1,2}\}$, 
which are also singlets of the rank 8 hidden sector gauge group. 
The $SO(10)$ singlet states $H^+$ and $H^-$ in table \ref{tablehi}
carry similar $U(1)_\zeta$ charges and transform under the 
hidden sector gauge group. 
These exotic states arise in the string models due to the 
breaking of the $E_6$ symmetry by Wilson lines. However, 
as the Wilson line is realised in the free fermionic construction
in terms of a GGSO phase, its precise identification is not 
a simple exercise. Its imprint is revealed due to the exotic 
charges, which will not have been generated otherwise. Furthermore, 
we note from table \ref{tableb} 
that the string model does contain the $SO(10)$ singlet states 
$S$ and $\overline S$,
with standard $E_6$ charges to break $U(1)_\zeta$ along flat 
directions. Therefore, this model can realise the Wilsonian
singlet dark matter scenario articulated in ref. \cite{ssr}. 
In the next section we turn to examine this question. 

\section{$SO(10)$ singlet Wilsonian dark matter }
\label{wildarmat}

In this section we examine several scenarios in which the Wilsonian 
matter states can account for the dark matter without overclosing 
the universe. Our analysis here is primarily qualitative and 
more detailed numerical analysis will be reported in future 
work. As discussed in the previous sections, 
the main feature of the Wilsonian matter states is the existence 
of an intrinsic stringy mechanism that produces stable matter 
states. In ref. \cite{ssr} a similar analysis was performed 
for states that carry Standard Model, or $SO(10)$ charges. 
The novelty here is that the Wilsonian states arise as $SO(10)$ 
singlets and interact with the Standard Model states only via 
the $U(1)_{Z^\prime}$ couplings. We note that contrary to other 
dark matter candidates in the literature, whose stability relies 
on the existence of global gauge or discrete symmetry, the stability 
of the Wilsonian states arises from the assumption that the Wilsonian
states themselves do not receive a vacuum expectation value. 
As seen from table \ref{tableb},
the heterotic--string model of ref. \cite{frzprime} allows this 
assumption to be made because it contains the standard $E_6$
charged stated $F_R$, ${\bar F}_R$ and $S$, $\bar S$ to break the 
gauge symmetry along flat directions. 

We next comment on the allowed values of the gauge and Yukawa 
couplings of the Wilsonian states. The entire cubic level 
superpotential was presented in ref. \cite{frzprime}. 
All the couplings in heterotic--string model are given in terms of the
unified gauge coupling and the VEVs of some moduli fields. The relevant
parameter for the calculation of the relic abundance is the $Z^\prime$ gauge 
coupling, subject to the assumptions on the $Z^\prime$ breaking scale, the 
mass scale of the dark matter states and the reheating temperature. 
These three scales are taken as input parameters and the constraints 
on the relic abundance are obtained subject to some initial assumptions
(i.e. thermal or non-thermal relic) by solving the Boltzmann equation. 
We consider both high and low scale $U(1)_{Z^\prime}$ breaking. 
In the low scale breaking scenario the $Z^\prime$ mass scale
can be generated dynamically, in which case the relevant parameters are
cubic level coupling in the string derived superpotential, which are
all given in terms of the unified gauge coupling. The numerical 
value of the gauge coupling at the unification scale is constrained 
by compatibility with the gauge coupling data at the electroweak scale. 
The $Z^\prime$ is constrained by the LHC experiments to be heavier than
a few TeVs and we may therefore assume that it is heavier than 
$\simeq 4$TeV.
In this case its mixing with the Standard Model $Z$--boson is 
small and does not affect the analysis. 
A detailed numerical analysis is beyond the scope of this paper, 
and will be reported in future publications. 

The low energy spectrum of the string model consists of the states 
in table \ref{table27rot}, where we also allow for the possibility of 
completely neutral states that may correspond to light hidden 
sector states. 
\begin{table}[!h]
\noindent 
{\small
\begin{center}
{\tabulinesep=1.2mm
\begin{tabu}{|l|cc|c|c|c|}
\hline
Field &$\hphantom{\times}SU(3)_C$&$ SU(2)_L $
&${U(1)}_{Y}$&${U(1)}_{Z^\prime}$  \\
\hline
$Q_L^i$&    $3$       &  $2$ &  $+\frac{1}{6}$   & $-\frac{2}{5}$   ~~  \\
$u_L^i$&    ${\bar3}$ &  $1$ &  $-\frac{2}{3}$   & $-\frac{2}{5}$   ~~  \\
$d_L^i$&    ${\bar3}$ &  $1$ &  $+\frac{1}{3}$   & $-\frac{4}{5}$  ~~  \\
$e_L^i$&    $1$       &  $1$ &  $+1          $   & $-\frac{2}{5}$  ~~  \\
$L_L^i$&    $1$       &  $2$ &  $-\frac{1}{2}$   & $-\frac{4}{5}$  ~~  \\
%
\hline
$D^i$       & $3$     & $1$ & $-\frac{1}{3}$     & $+\frac{4}{5}$  ~~    \\
${\bar D}^i$& ${\bar3}$ & $1$ &  $+\frac{1}{3}$  &   $+\frac{6}{5}$  ~~    \\
$H^i$       & $1$       & $2$ &  $-\frac{1}{2}$   &  $+\frac{6}{5}$ ~~    \\
${\bar H}^i$& $1$       & $2$ &  $+\frac{1}{2}$   &   $+\frac{4}{5}$   ~~  \\
\hline
$S^i$       & $1$       & $1$ &  ~~$0$  &  $-2$       ~~   \\
%
%
\hline
${\cal D}$  & $3$       & $1$ &  $-\frac{1}{3}$  &  $+\frac{4}{5}$  ~~    \\
${\bar{\cal D}}$&${\bar 3}$& $1$ &  $+\frac{1}{3}$  &  $-\frac{4}{5}$  ~~ \\
\hline
$\phi$       & $1$       & $1$ &  ~~$0$         & $-1$     ~~   \\
$\bar\phi$       & $1$       & $1$ &  ~~$0$     & $+1$     ~~   \\
\hline
\hline
$\zeta^i$       & $1$       & $1$ &  ~~$0$  &  ~~$0$       ~~   \\
\hline
\end{tabu}}
\end{center}
}
\caption{\label{table27rot}
\it
Spectrum and
$SU(3)_C\times SU(2)_L\times U(1)_{Y}\times U(1)_{{Z}^\prime}$ 
quantum numbers, with $i=1,2,3$ for the three light 
generations. The charges are displayed in the 
normalisation used in free fermionic 
heterotic--string models. }
\end{table}

The trilinear superpotential embedding the supersymmetric
$Z^\prime$ model and the relevant interactions of the $\phi$ and $\bar \phi$ is
\bea
\mathcal{W} &=&
Y^u{}_{i,j,k} \bar{u}_R^i\,Q_L^j\,\bar{H}^k - 
Y^d{}_{i,j,k} \bar{d}_R^i\,Q_L^j\,H^k - 
Y^e{}_{i,j,k} \bar{e}_R^i\,L_L^j\,H^k \nn \\
&+& \lambda _{i,j,k}\, S^i\,H^j\,\bar{H}^k + 
\kappa _{i,j,k}\, S^i\,\bar{D}^j\,D^k +  
\lambda ^s{}_i S^i\,\bar{\phi}\,\bar{\phi} 
~.
\label{superpot}
\eea
As seen from eq. (\ref{superpot}) there are
no terms that allow for the $\phi, \bar\phi$ states to decay 
to lighter states at leading order. Breaking the $U(1)_{Z^\prime}$ 
symmetry with the VEV of $S^i$ leaves a remnant discrete symmetry 
\cite{lds, ssr} which forbids their decay at any order in the
superpotential. A potential mass term for $\bar\phi$ 
arises in the trilinear superpotential, eq. (\ref{superpot}). 
Additional mass terms, that are invariant under all gauge and discrete
symmetries, may be generated from higher order terms. Therefore, 
the Wilsonian matter states, namely the fermionic component
arising from the $\phi, \bar \phi$ states, in the string
derived $Z^\prime$ model
are heavy and stable. Their mass density may overclose the 
universe if they are over abundant. We refer to these states 
as Wilsonian singlet dark matter, or $W_s$ for short. 

The Wilsonian singlet interacts with the 
non--exotic states in table \ref{table27rot}
only via the $Z^\prime$ gauge charges. 
Its number density can change only by annihilations 
via the diagrams in figures \ref{wswsff} and \ref{wswsfsfs}
into fermions and their superpartners, and, 
depending on the $U(1)_{Z^\prime}$ symmetry breaking scale, 
into the gauge bosons and their superpartners, as in figures 
\ref{wswsZZTU} and \ref{wswsZsZsTU}. 

\begin{center}
\begin{figure}[h]
\centering
\includegraphics[width=5cm, angle=0]{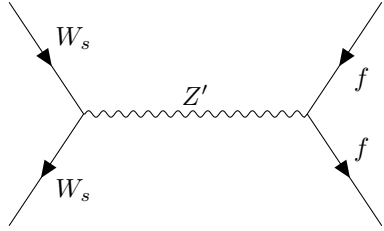}
 \caption{\it $W_s\bar{W}_s\longrightarrow f\bar{f}$ decay.}
\label{wswsff}
\end{figure}

\begin{figure}[h]
\centering
\includegraphics[width=5cm, angle=0]{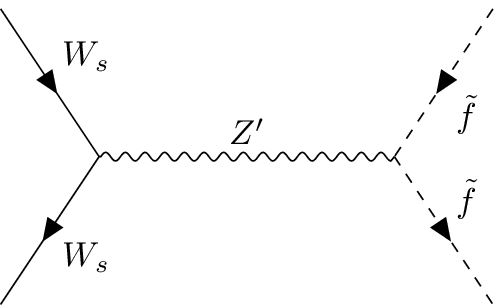}
 \caption{\it $W_s\bar{W}_s\longrightarrow \tilde{f}\tilde{f}^*$ decay.}
\label{wswsfsfs}
\end{figure}
\end{center}

The annihilation into fermion and sfermion in figures 
\ref{wswsff} and \ref{wswsfsfs}
leads to the two contributions
\bea
\sigma_{W_sW_s\rightarrow ff} = \frac{4\,\pi}{3}\frac{N_{Z'} \sqrt{S} \,
\left(2\,M^2_{W_s} + S\right)}{\sqrt{S - 4\,M^2_{W_s}}\left(M^2_{Z'} -
S\right)^2}
\eea
and
\bea \label{sigmascalar}
\sigma_{W_sW_s\rightarrow \tilde{f}\tilde{f}} = \frac{\pi}{3}\frac{N_{Z'}
\sqrt{S} \, \left(2\,M^2_{W_s} + S\right)}{\sqrt{S -
4\,M^2_{W_s}}\left(M^2_{Z'} - S\right)^2}
\eea
where we defined the parameter
\bea
N_{Z'} = \frac{g'^4}{16\,\pi^2}\, Q^2_f \, Q^2_{W_s}
\eea
with $Q_i$ and $g'$ being the $U(1)_{Z'}$ charge and the corresponding coupling constant, respectively. 
The $N_{Z'}$ parameter accounts for the strength of a $Z'$ exchange. 
The total cross-section for the
annihilation into fermions and sfermions
$W_s \bar{W}_s \rightarrow f \bar{f}, \tilde{f_L} \bar{\tilde{f_L}},
\tilde{f_R} \bar{\tilde{f_R}}$
is now easily reached
\bea \label{TotFF}
\sigma_{W_sW_s\rightarrow ff,\tilde{f}\tilde{f}}  = \sigma_{W_sW_s\rightarrow
ff} + 2\,\sigma_{W_sW_s\rightarrow \tilde{f}\tilde{f}} =  \frac{2 \pi N_{Z'}
\sqrt{S} \, \left(2\,M^2_{W_s} + S\right)}{\sqrt{S -
4\,M^2_{W_s}}\left(M^2_{Z'} - S\right)^2}
\eea
where the contribution of (\ref{sigmascalar}) has been doubled to account for
the separate events
$W_s \bar{W}_s \rightarrow \tilde{f_L} \bar{\tilde{f_L}}$ and $W_s \bar{W}_s
\rightarrow \tilde{f_R} \bar{\tilde{f_R}}$
which have identical cross-sections.
\begin{center}
\begin{figure}[h]
\centering
\includegraphics[width=10cm, angle=0]{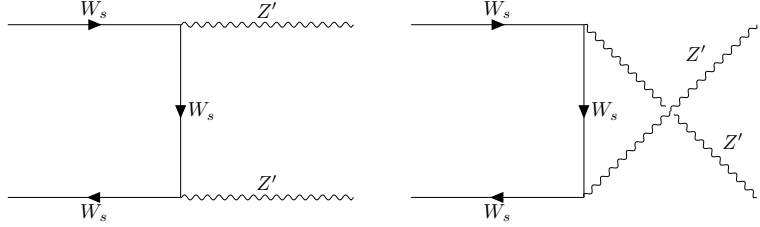}
 \caption{\it  $W_s\bar{W}_s\longrightarrow Z' Z'$ decay.}
\label{wswsZZTU}
\end{figure}
\begin{figure}[h]
\centering
\includegraphics[width=10cm, angle=0]{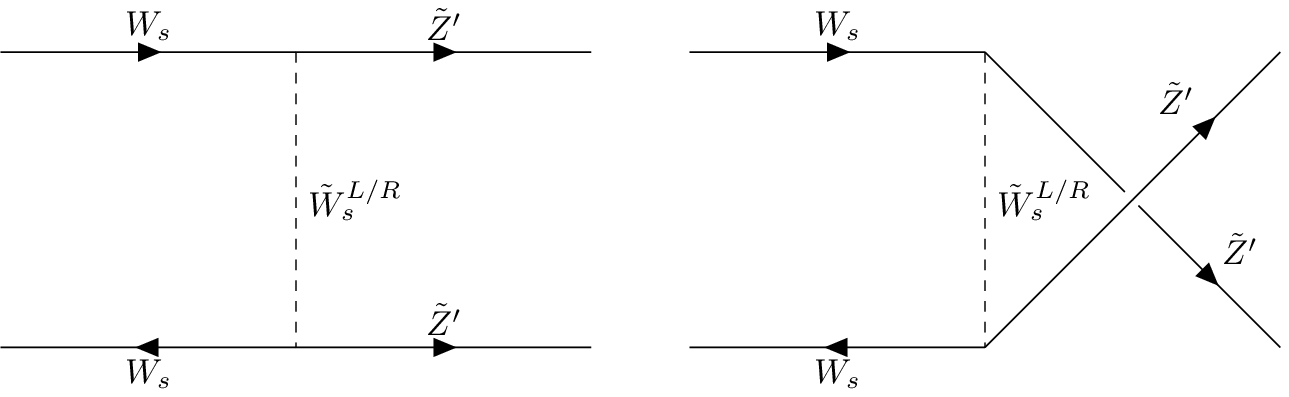}
 \caption{\it  $W_s\bar{W}_s\longrightarrow \tilde{Z}'\tilde{Z}'$ decay.}
\label{wswsZsZsTU}
\end{figure}
\end{center}

The computation of the annihilations into vector bosons in figure
\ref{wswsZZTU} and into their superpartners in figure \ref{wswsZsZsTU}
leads to the cross-sections
\beqn \label{ZZ}
\sigma_{W_sW_s\rightarrow Z'Z'} =  4\,\pi\,\tilde{N}_{Z'} \frac{\left(S^2 +
4 M^2_{W_s}\,S - 8\,M^4_{W_s}\right)\ln\left(\frac{S + A}{S-A}\right) - A
\left(S + 4 M^2_{W_s} \right)}{A^2 S}
\eeqn
and
\bea \label{ZfZf}
\sigma_{W_sW_s\rightarrow \tilde{Z}'\tilde{Z}'} =  2\,\pi\,\tilde{N}_{Z'}
\frac{A - 2 M^2_{W_s}\ln\left(\frac{S+A}{S-A}\right)}{A^2}
\eea
where we have defined $\tilde{N}_{Z'}$ as
\bea
\tilde{N}_{Z'} = \frac{g'^4}{16\,\pi^2}\,  Q^4_{W_s}
\eea
and the kinematical parameter $A = \sqrt{S (S - 4 M^2_{W_s})}$.
Since $W_s$ is heavy and stable, its mass density is constrained 
by the requirement that it does not overclose the universe. 
Alternatively, we may extract the regions of parameter 
space where the non--baryonic dark matter abundance 
can be explained in terms of the Wilsonian singlet dark matter. 
After $U(1)_{Z^\prime}$ symmetry breaking the $W_s$ interactions
are suppressed by $1/M_{Z^\prime}^2$ and it can be classified 
as weakly interacting massive particle.
It decouples from the thermal 
bath when its annihilation rate falls behind the expansion 
rate of the universe.
The annihilation rate of a particle is
\begin{eqnarray}
\Gamma =
 \langle\sigma_{\rm ann}|v|\rangle n_{_{EQ}},
\end{eqnarray}
where the number density at the equilibrium, $n_{_{EQ}}$, 
is given by
\begin{equation}
{\displaystyle{n_{_{EQ}}}}=
\begin{cases}
{\displaystyle{g_{\rm eff}\left(\frac{\zeta(3)}{\pi^2}\right)T^3}}&
			$\hbox{~~~relativistic}$\cr
{\displaystyle{g_{\rm eff}\left(\frac{mT}{2\pi}\right)^{3/2}\exp(-M/T)}}&
	      		$\hbox{~~~non-relativistic}$\cr
\end{cases}
\end{equation}
and $\zeta(3)=1.20206$ is the Riemann zeta function of 3,
and $g_{\rm eff}$ is the effective number of
degrees of freedom of the particle.
In the expanding universe the evolution equation of 
the number density is described by the Boltzmann equation
\cite{Kolb:1990vq}. 
In terms of the number density in a comoving volume
$Y=n/s_{\rm e}$, where $s_{\rm e}$ is the entropy density,
and the dimensionless parameter $x=M/T$, the 
Boltzmann equation takes the form
\bea \label{Boltzmann}
\frac{d\,Y}{d\, x} = - \lambda x^{-2}\,\left(Y^2 - Y^2_{EQ}\right)\, ,
\eea
with $\lambda$ related to the interaction rate of the particle through
\bea \label{lambda}
\lambda = 0.26 <\sigma_{ann} |v|> M_{W_s} m_{Pl} 
            \frac{g_{*s}}{\sqrt{g_{*}}} \,
\eea
with $Y_{EQ}$ the density at thermal equilibrium and $m_{Pl}$ the Planck Mass.
In (\ref{lambda}) the variables $g_*$ and $g_{*s}$ count, for different
purposes, the effective relativistic degrees of freedom at a given $T$
\bea
g_{*} &=& \sum_{i=bosons}\,g_i \left(\frac{T_i}{T}\right)^4 +
\frac{7}{8}\,\sum_{i=fermions}\,g_i \left(\frac{T_i}{T}\right)^4\, , \nn \\
g_{*s} &=& \sum_{i=bosons}\,g_i \left(\frac{T_i}{T}\right)^3 +
\frac{7}{8}\,\sum_{i=fermions}\,g_i \left(\frac{T_i}{T}\right)^3 \, ,
\eea
where $g_i$ are the complete internal degrees of freedom and $T_i$ the
temperature of the given particle \emph{i}.
The annihilation cross-section $\sigma_{ann}$ determines the evolution of the
relic density via its thermal average, $<\sigma_{ann} |v|>$, 
and its computation must be performed to calculate 
the relic abundance left by $W_s$.
We can distinguish different scenarios for the Wilsonian 
singlet dark matter, depending on the $Z^\prime$ 
symmetry breaking scale, $M_{Z^\prime}$, the mass scale of 
$W_s$, $M_{W_s}\equiv M$, and the reheating temperature, $T_R$, 
following inflation. The $Z^\prime$ breaking scale can vary 
from the experimental LHC mass limits, of the order of 
a few TeV, up to the Planck scale.
The constraints on the $W_s$ mass can vary from being ultra light \cite{hotw}
to being super--massive \cite{ssr, ckr}, depending 
on the $Z^\prime$ breaking scale, and the reheating temperature.
There are several possible dark matter scenarios 
for the Wilsonian singlet $W_s$: 

\begin{itemize}

\item {\bf $M_{W_s}>>M_{Z^\prime}$ without inflation. }
In this case the Wilsonian singlet is strongly interacting in the early 
universe and remains in thermal equilibrium 
until it becomes non--relativistic. 
The
solution of the Boltzmann equation leads to a density value of
\bea \label{Y0nonrel}
Y_0 =
\frac{3.79}{m_{Pl}\,T_{dec}\,<\sigma_{ann}|v|>}\,
\left(\frac{\sqrt{g_*}}{g_{*s}}\right).
\eea
In this scenario all the annhilation channels depicted 
in figures \ref{wswsff}, \ref{wswsfsfs}, \ref{wswsZZTU}, \ref{wswsZsZsTU}
are open. The $s$-wave expansion of the Boltzmann equation 
\cite{Kolb:1990vq} yields the thermal average of (\ref{TotFF}), (\ref{ZZ})
and (\ref{ZfZf}) as: 
\bea \label{relFFSS}
<\sigma_{W_sW_s\rightarrow ff,\tilde{f}\tilde{f}}|v|>  = \frac{8 \pi
N_{Z'}}{3\,M_{W_s}^2} + O(v^2)
\eea
and 
\bea
<\sigma_{W_sW_s\rightarrow Z'Z'}|v|> + <\sigma_{W_sW_s\rightarrow
\tilde{Z}'\tilde{Z}'}|v|> =  \frac{4 \pi \tilde{N}_{Z'}}{M_{W_s}^2} + O(v^2)\,
\eea
which sums with (\ref{relFFSS}) for the total annihilation cross-section of the
singlet in the regime $M_{W_{s}} >> M_{Z'}$
\bea \label{rel2}
<\sigma_{ann}|v|> = \sigma_0 + O(v^2) = \frac{4
\pi}{M_{W_s}^2}\left(\frac{2}{3}\,N_{Z'} + \tilde{N}_{Z'} \right) + O(v^2)\, \,
{}.
\eea
To complete the computation of the number density (\ref{Y0nonrel}) the
decoupling temperature is inferred by exploiting, from the Boltzmann
equation, the condition $d Y/d x \simeq 0$ which results in \cite{Kolb:1990vq}
\beqn
& &\frac{T_{dec}}{M_{W_s}} \simeq   \nonumber\\
& &\left(\ln\left(0.038\,(g/\sqrt{g_{*}})\,m_{Pl}
M_{W_s} \sigma_{0}\right)-
 \frac{1}{2}\,\ln\left(\ln\left(0.038\,(g/\sqrt{g_{*}})\,m_{Pl} M_{W_s}
\sigma_{0}\right)\right) \right)^{-1} .~~~~~\nonumber 
\eeqn
Its easy to show that the decoupling temperature has a very mild
dependence over the model dependent factors $N_{Z'}$ and $\tilde{N}_{Z'}$,
as well as over realistic values of the variable $g_{*}$. 
A solid and simple fit
is found using the formula \cite{ssr}
\bea
T_{dec} = \frac{M_{W_{s}}}{\ln\left(m_{Pl}/M_{W_s}\right)}\,,
\eea
which allows to find the relation for the density $Y_0$ in
eq. (\ref{Y0nonrel})
\bea
Y_0 \simeq \frac{0.3 M_{W_s}
\ln(m_{Pl}/M_{W_s})}{\sqrt{g_*}\,m_{Pl}\,(\frac{2}{3}\,N_{Z'} + \tilde{N}_{Z'}
)}
\eea
where the approximation $g_* \sim g_{*s}$ can be used as the decoupling
temperature is sufficiently high in the relevant region of parameter space.
Finally, to draw an estimate for the value of $\Omega\,h^2$, we first find the
current energy density $\rho_0$ by multiplying for the entropy $s_{e_0}$,
 $\rho_0 = s_{e0}\,Y_0\,M$, and then divide by the critical energy $\rho_c$ to
arrive at the final expression
\bea \label{Bound1}
\Omega\,h^2 = \frac{\rho_0\,h^2}{\rho_c} = \frac{2970\,
M_{W_s}\,Y_0\,cm^{-3}}{1.05\,10^4\,eV\,cm^{-3}}\, .
\eea
Taking $\Omega h^2 \sim 1$ we obtain an upper bound
\bea \label{Bound}
M_{W_s} < 10^{5}\,\sqrt{\frac{\sqrt{g_*}\,
\mathcal{N}}{\ln(m_{Pl}/M_{W_s})}} \GeV
\eea
with ${\mathcal{N}} = \left({2\over3}\,N_{Z'} + \tilde{N}_{Z'} \right)$
determined by the $U(1)_{Z^\prime}$ couplings in table \ref{table27rot}. 
We note that $U(1)_{Z^\prime}$ symmetry is restored when the temperature
goes above its breaking scale. The extra vector--like matter states,
beyond the Standard Model, in table \ref{table27rot}, become massive
at that scale. Their effect above that scale is incorporated in the 
analysis by summing over the charges in the numerical factor under the 
square root in eq. (\ref{Bound}).
By an explicit display of the bound in eq. (\ref{Bound})
we may estimate the values of $g'$ and $M_{W_s}$ that are required, 
in such a scenario, 
to avoid an over abundant Wilsonian state relic. 
The region allowed by (\ref{Bound}) in figure \ref{RelicBound} reveals how,
to avoid too light $Z'$
mass and simultaneously respect the hierarchy $M_{W_s}>>M_{Z^\prime}$,
a large, but perturbative, $g^\prime$ must be adopted. 
We notice that perturbativity of the $U(1)_{Z^\prime}$ coupling
requires an upper bound of
$\simeq 25$TeV on the mass of the Wilsonian dark matter as thermal 
relic with low scale $U(1)_{Z^\prime}$ breaking.
\begin{center}
\begin{figure}[h]
	\centering
	\includegraphics[width=10cm, angle=0]{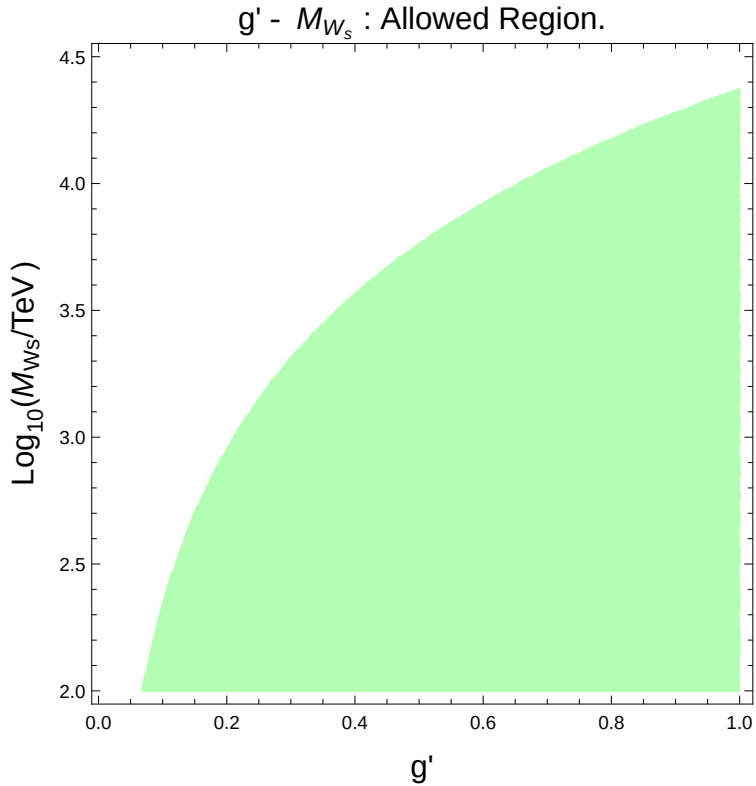}
	\caption{\it  In green: region in the $g^\prime-M_{W_s}$ space preventing overabundant Dark Matter
	in the case $M_{W_s}>>M_{Z^\prime}$ without inflation.}
	\label{RelicBound}
\end{figure}
\end{center}


\item {\bf $M_{W_s}>>M_{Z^\prime}$ with inflation. }
The current relic density could also be generated by an out-of-equilibrium
production after reheating if such process takes place after singlet
decoupling.
In this case inflation will dilute the singlet density and, from the Boltzmann
equations (\ref{Boltzmann}), we can study the subsequent evolution with the
approximate knowledge of the density derivative,
\bea \label{BoltzmannZero}
\frac{d\,Y}{d\, x} =  \lambda x^{-2}\,Y^2_{EQ}\, ,
\eea
with $Y_{EQ} = 0.145 g_{eff}/g_{*s} \,x^{3/2}\,e^{-x}$ (non relativistic case).
When $\lambda$ is independent of $x$, eq. (\ref{BoltzmannZero}) can be promptly
integrated to the present temperature, providing the density produced by the
singlet
after reheating \cite{ssr}
\bea \label{YOInflaton}
Y_0 = \frac{\lambda g_{eff}^2}{2} \left(\frac{0.145}{g_{*}} \right)^2
\left(\frac{M_{W_s}}{T_R} + \frac{1}{2}\right)\, e^{-2\,\frac{M_{W_s}}{T_R}}\,
,
\eea
with $T_R$ the reheating temperature.
Such value, inserted in (\ref{Bound1}) and requiring that it does not 
exceed the measured cosmological relic density, results in a bound, 
involving the $Z'$ and $W_{s}$ masses as well as $T_R$, whose specific form depends on the process 
through which the final density is regenerated.
As long as  the reheating temperature is greater than the $Z'$ mass,
the generation of the singlets will take place with all the channels
investigated above, so that
the relevant (thermal) cross-section is (\ref{rel2}). 
Using (\ref{YOInflaton}),
the cosmological upper bound over the relic is obtained with the condition
\bea
M_{W_s} > T_R \left(26 + \frac{1}{2} \ln{\left(\frac{M_{W_s}}{T_R} +
\frac{1}{2} \right)} + \frac{1}{2} \ln{\left( \mathcal{N} \right)} \right)
\eea
and ${\mathcal{N}} = \left(\frac{2}{3}\,N_{Z'} + \tilde{N}_{Z'} \right)$.

\item {\bf $M_{W_s}<<M_{Z^\prime}$ without inflation. }
In this case $W_s$ is a WIMP and it can only annihilate into the 
matter supermultiplets in table \ref{table27rot} via the 
diagrams in figures \ref{wswsff} and \ref{wswsfsfs}, 
which are suppressed by the heavy $Z^\prime$ vector boson mass.
The interaction between $W_s$  
and other matter states will be kinematically suppressed 
when $T$ is below $M_{Z'}$. Decoupling therefore 
occurs when $W_s$ is still relativistic at freeze-out. 
The resulting
density in the regime $T \gg M$ can be analytically extracted from the
Boltzmann equation \cite{Kolb:1990vq}
\bea \label{Y0}
Y_0 = 0.2\,\frac{g_{{}_{Ws}}}{g_{*s\left(T_{dec}\right)}}  \, .
\eea
To obtain an estimate for the value of $\Omega\,h^2$ we use the 
expression in eq. (\ref{Bound1}).
We first find the current
energy density 
 $\rho_0 = s_{e0}\,Y_0\,M$, and then divide by the critical energy $\rho_c$ to
arrive at the expression in eq. (\ref{Bound1}).
The resulting formula can then be used to obtain a constraint on $M_{W_s}$
\bea
M_{W_s} < \frac{\Omega\,h^2\,3.5\,eV}{Y_0} \simeq \left(17.7 eV\right)
\,\frac{g_{*s\left(T_{dec}\right)}}{g_{{}_{Ws}}}\,\Omega\,h^2\, .
\eea
If we consider a scenario with $M_{Z'}$ at the TeV scale, with a similar range
for the decoupling temperature, the number of the degrees of freedom still
relativistic will be of order $10^2$. 
Taking $\Omega\, h^2\sim 1$ 
we obtain the limit
\bea
M_{W_s} \lsim 1 keV \, .
\label{mwsltmzprimebound}
\eea
Such tight constraint is typical of over abundant relativistic 
WIMP particle, as the condition $M_{W_s} < M_{Z'}$ forces the 
singlet to be light because of the suppression of the interactions
by a factor $1/M_{Z'}$. 

\item {\bf $M_{W_s}<<M_{Z^\prime}$ with inflation. }

The constraint in eq. (\ref{mwsltmzprimebound}) is relaxed in the 
presence of inflation. In this case $W_s$ is completely diluted 
by inflation and is regenerated after reheating. 
In the limit $M<M_{Z^\prime}$ and $T_R<M_{Z^\prime}$, 
we approximate the $Z^\prime$--mediated interaction by a four--point
Fermi interaction.
In this case $W_s$ can only annihilate via the diagrams 
in figures \ref{wswsff} and \ref{wswsfsfs} into 
matter states, but not those in figures \ref{wswsZZTU} and 
\ref{wswsZsZsTU} into gauge bosons and gauginos.
The thermal cross-section in the non-relativistic limit $T_R < M_{W_s}$
is given by 
\bea
\sigma_0  = \frac{32 \pi\,M_{W_s}^2}{M_{Z'}^4}\,N_{Z'}\, \, ,
\eea
and by 
\bea
\sigma_0  = \frac{2 \pi\,s}{M_{Z'}^4}\,N_{Z'}\, \, .
\eea
in the relativistic limit $T_R > M_{W_s}$. 
The non--relativistic case is a replication of the analysis 
for the non-relativistic, out-of-equilibrium production $W_s$ 
when $T_R > M_{Z^\prime}$. 
The resulting bound is
\bea
M_{W_s} > T_R \left(27 + \frac{1}{2} \ln{\left(\frac{M^5_{W_s}}{M_{Z'}^4\,T_R}
+ \frac{M_{W_s}^4}{2\,M_{Z'}^4} \right)} + \frac{1}{2} \ln{\left( N_{Z'}
\right)} \right) \, .
\eea
In the relativistic limit, the thermal cross section 
defines a temperature-dependent $\lambda$
parameter and the integration of the Boltzmann equations
requires more care. 
To proceed we can express the Center of Mass energy of the process
as a thermal average, yielding its dependence on the temperature as
\bea
<s> = 4 <E^2> \simeq \left(\frac{5}{4}\right)\,40\,T^2\, .
\eea
The integration of (\ref{BoltzmannZero}), with a relativistic equilibrium
density, is now at hand and the final constraint on the Wilsonian 
singlet mass bound takes the form
\bea
\frac{M_{W_s}\,T_R^3}{M_{Z'}^4} < 1.6 \times
10^{-28}\,\left(\frac{g_{*}^{3/2}}{N_{Z'}\,g_{eff}^2}\right)\, .
\label{mwstrmzprime}
\eea
As there are three unknown parameters ($M_{W_s},\, M_{Z'},\, T_R$)
in Eq.~(\ref{mwstrmzprime}) we cannot
infer a definite value for the mass of $W_s$ and $Z'$.
We may nevertheless conclude that $Z'$ should be very heavy.

\end{itemize}

Finally we remark that direct or indirect dark matter detection of the
Wilsonian dark matter candidates will be extremely challenging, 
as those are expected to be weaker than the prevailing constraints
on, say, neutralinos. The reason being that the interaction of the
Wilsonian matter states with the Standard Model particles 
is governed by the $Z^\prime$ mass scale, 
which is higher compared to the weak scale that typifies the 
neutralino interactions. Similarly, indirect detection is suppressed 
due to the low trapping rate of the Wilsonian singlet matter states
in, {\it e.g.}, the sun \cite{fop}. 

\section{Conclusions}\label{concusion}

The issue of dark matter is one of the important enigmas of 
modern physics. The problem is exacerbated as plausible  
particle candidates can vary from the ultra--light \cite{hotw}
to the super--massive \cite{ssr, ckr}. It is then  sensible to 
seek guidance from string theory, which is the 
only contemporary  approach that consistently unifies 
gravity with the gauge interactions. 

We emphasise that proposals of dark matter candidates 
are ample in the literature without reference to 
their gravitational or string theory connections. 
Our approach aims to incorporate 
constraints from string theory, which is a contemporary 
framework compatible with quantum gravity, in the analysis
In that respect we note that, for example,
since the mid eighties there is a plethora of string 
inspired $Z^\prime$ studies, whereas, to date, the only known worldsheet 
construction that allows for an unbroken $U(1)_{Z^\prime}$ symmetry of the
type discussed in the string inspired literature is that of 
ref. \cite{frzprime}. 
Wilsonian matter states are a generic consequence of symmetry breaking 
in string theory by Wilson lines. The main feature which characterises 
them is the existence of an intrinsic string stabilisation mechanism 
that forbids their decay to the standard model states. This intrinsic 
stabilisation mechanism provides the main motivation to study them. 

It is then amply rewarding that string constructions 
indeed contain in them the intrinsic ingredients to 
produce heavy and stable dark matter. 
The stabilisation mechanism arises due to the
breaking of non--Abelian gauge symmetries 
in string theory by Wilson line, which gives rise
to exotic states that do not satisfy the 
quantisation conditions of the unbroken
non--Abelian gauge symmetry. Well known 
examples of such states include those that
carry fractional electric charge that are 
highly constrained by experiments. The favoured class 
of Wilsonian states that can constitute the 
dark matter are those that are neutral under
the Standard Model, but carry fractional charge 
under an extra $U(1)_{Z^\prime}$ symmetry. While 
this possibility has been entertained before 
in \cite{ssr}, the string derived $Z^\prime$ model 
of ref. \cite{frzprime} is the first concrete
string derived model in which the Wilsonian
singlet dark matter can be realised. This model
contains all the ingredients needed to break the 
$U(1)_{Z^\prime}$ symmetry at a high or low scale,
while maintaining a discrete symmetry that forbids the
decay of the Wilsonian singlet matter state. 
The Wilsonian dark matter singlets in the model
of ref. \cite{frzprime} are $SO(10)$ singlets 
and arise from the breaking of $E_6\rightarrow SO(10)\times U(1)$ 
by Wilson lines. 
Even within the constraints of the string 
construction, as we discussed in section \ref{wildarmat}, 
there are varied possibilities depending on
the $U(1)_{Z^\prime}$ breaking scale, the reheating temperature
$T_R$ and the mass of $W_s$ itself. 
We can then all but hope that forthcoming experiments 
will narrow down the possibilities by, for example, 
discovering a extra vector boson $Z^\prime$ in the 
vicinity of the multi--TeV scale.

\section*{Acknowledgments}

AEF thanks the theoretical physics department at Oxford
University for hospitality.
AEF is supported in part by the STFC (ST/L000431/1). 
LDR is supported by the STFC/COFUND Rutherford International Fellowship scheme.
The work of CM is supported by the “Angelo Della Riccia” foundation and by 
the Centre of Excellence project No TK133 “Dark Side of the Universe”.

\end{document}